\documentclass{getwriting}
\usepackage{amsfonts, amsmath, amssymb}
\usepackage{overpic}
\usepackage{multirow}
\usepackage[sfdefault]{atkinson}
\usepackage{tikz}
\usetikzlibrary{arrows.meta, calc}
\title{The role of estrogen receptor $\alpha$ on calcium transport during smooth muscle contractions}
\author[1]{Rebecca M Crossley \orcidlink{0000-0001-7342-0207} \thanks{crossley@maths.ox.ac.uk}}
\author[2]{Jessica R Crawshaw \thanks{jess.crawshaw@qut.edu.au}}
\author[3]{Neda K Joniani \thanks{n.khodabakhsh@sydney.edu.au}}
\author[4]{Ellen~T~Kahiya~\thanks{kahiyae@faf.cuni.cz}}
\author[5]{Lata I Paea \thanks{lpae237@aucklanduni.ac.nz}}
\author[5]{Alys R Clark \thanks{alys.clark@auckland.ac.nz}}
\affil[1]{\footnotesize Mathematical Institute, University of Oxford, Oxford, United Kingdom}
\affil[2]{\footnotesize Queensland University of Technology, Brisbane, Australia}
\affil[3]{\footnotesize University of Sydney, Sydney, Australia}
\affil[4]{\footnotesize Charles University, Prague, Czech Republic}
\affil[5]{\footnotesize University of Auckland, Auckland, New Zealand}

\newcommand{\Jin}{$J_{\mathrm{in}}$}
\newcommand{\Jserca}{$J_{\mathrm{SERCA}}$}
\newcommand{\Jleak}{$J_{\mathrm{leak}}$}
\newcommand{\JIPthreeR}{$J_{\mathrm{IP_3R}}$}
\newcommand{\Jryr}{$J_{\mathrm{RyR}}$}
\newcommand{\Jpmca}{$J_{\mathrm{PMCA}}$}

\date{}

\begin{document}
\maketitle
\vspace{-.8cm}
\begin{center}


\end{center}

\vspace{.4cm}

\begin{abstract}
Reproductive hormones regulate a wide range of physiological processes throughout the human lifespan. 
Estrogen, in particular, varies substantially across the menstrual cycle and is widely used in contraceptives and hormone replacement therapies. 
Despite its physiological importance, few experimental studies and even fewer mathematical models explicitly investigate how estrogen regulates smooth muscle function. As smooth muscle lines our blood vessels, airways, uterus, and several other organs, understanding how estrogen impacts smooth muscle is important to improving the understanding of sex differences in lifelong health. Here we extend an established mathematical model of smooth muscle cell calcium signalling to incorporate estrogen-dependent modulation of intracellular calcium transport pathways. 
Numerical simulations, global sensitivity analysis, and numerical bifurcation analysis are then used to quantify the influence of estrogen on intracellular calcium dynamics and the resulting steady-state and oscillatory behaviours. 
Our results demonstrate that physiologically relevant changes in estrogen shift intracellular calcium concentrations while leaving the underlying bifurcation structure and qualitative dynamics largely unchanged, suggesting that estrogen acts primarily as a quantitative modulator of smooth muscle calcium signalling. This work also serves to establish a foundation for future mechanistic models of hormone-dependent cell physiology.
\end{abstract}
\section{Introduction}
\label{sec:intro}

Sex and gender have historically been excluded as biological variables in research studies, despite being important factors that influence the timing of onset and outcomes experienced in a number of medical conditions~\cite{usselman2024}. 
Estrogen is an example of just one key reproductive hormone whose circulating concentration varies substantially throughout the menstrual cycle, pregnancy, and the menopause, and may also be modified through external factors such as the use of hormonal contraceptives or hormone replacement therapy~\cite{carmina}. 
Estrogen plays a central role in reproduction, and also regulates a wide range of physiological processes through its interactions with estrogen receptors expressed throughout the body. 
Among its many targets are smooth muscle tissues, where estrogen signalling influences vascular tone, airway function, uterine contractility, gastrointestinal motility, and lower urinary tract function~\cite{harvey2024sex, hilleubanks2011, senthilkumar2023, tran2020, xia2025}. 
Although these tissues perform diverse physiological roles, they all contain smooth muscle, which depends on a tightly regulated intracellular Ca$^{2+}$ signalling to control contraction and relaxation~\cite{karaki, somlyo, wray}. 
Consequently, understanding how estrogen regulates smooth muscle Ca$^{2+}$ dynamics is important for understanding both normal physiology and the development of hormone-dependent disease~\cite{bhallamudi2020}.

The cardiovascular system provides one of the clearest examples of estrogen's physiological importance. 
Before menopause, women have a lower incidence of hypertension and cardiovascular disease than age-matched men, but this advantage is rapidly lost, often irrevocably, following the menopausal decline in ovarian estrogen production~\cite{fasero2025, joyner2016, xia2025}. 
Experimental evidence has identified estrogen as a key protective factor, acting in part through its regulation of vascular smooth muscle Ca$^{2+}$ signalling~\cite{fasero2025, tran2020}. 
Estrogen has been shown to reduce cytosolic Ca$^{2+}$ availability through several complementary mechanisms, including suppression of voltage-dependent Ca$^{2+}$ channel activity and activation of large-conductance Ca$^{2+}$-activated potassium channels~\cite{asuncionalvarez2024, chen1999, han2006, han1995, hill2017, nakajima, tran2020, valverde1999, yu2011}. 
By reducing the intracellular Ca$^{2+}$ available to drive contraction, estrogen lowers vascular smooth muscle tone, contributing to reduced vascular resistance and blood pressure. 
When estrogen concentrations decline following menopause, this restraint on smooth muscle contractility is progressively lost, potentially contributing to increased cardiovascular risk observed in postmenopausal women~\cite{fasero2025}. When estrogen is delivered via hormone replacement therapy, its impact is complex, with current evidence suggesting that cardiovascular outcomes depend strongly on the timing of treatment initiation, with therapy begun near the onset of menopause generally associated with more favourable outcomes than therapy commenced years later~\cite{anagnostis2022, dcosta2025}. This suggests that both the current circulating concentration of estrogen, and prior exposure to estrogen  may be important to outcomes~\cite{nicholson2017, senthilkumar2023, xia2025}.  Understanding these changes in more detail requires a quantitative understanding of how estrogen regulates smooth muscle calcium signalling under normal physiological conditions.

The effects of estrogen on smooth muscle are mediated through estrogen receptors, particularly estrogen receptor $\alpha$ (ER$\alpha$), estrogen receptor $\beta$ (ER$\beta$), and G protein-coupled estrogen receptor (GPER)~\cite{nita2021}, with ER$\alpha$ being of particular interest in this study. Muscle cells contract via interaction between actin and myosin filaments within the cell, with calcium playing a key role in exposing actin to myosin allowing for contraction~\cite{hilleubanks2011}. Calcium transport across the plasma membrane and from intracellular stores into the cytosol is therefore essential for normal smooth muscle contractility~\cite{hilleubanks2011}. ER$\alpha$ may influence smooth muscle contraction by modulating calcium channels, pumps, and other transport proteins involved in calcium homeostasis~\cite{tran2020}. In addition, there are likely indirect effects on smooth muscle contraction related to ER$\alpha$ via nitric oxide signalling from endothelial cells~\cite{Traupe2007}. Experimental studies showing the effects of estrogen on calcium transport pathways allow mechanisms to be hpothesised, but as these mechanisms interact across multiple signalling pathways and timescales  it difficult to predict their combined influence on intracellular Ca$^{2+}$ dynamics. 
Mathematical modelling provides a complementary framework in which these interacting mechanisms can be integrated and their system-level consequences explored. Despite extensive mathematical models of calcium signalling and smooth muscle physiology~\cite{atri, goldbeter1990, jacob2007, koen2004, yang2003, Yang2006}, hormonal regulation, and sex differences more generally, are rarely incorporated into existing models~\cite{layton2021, usselman2024}. 
Consequently there remains no mathematical framework for systematically investigating how estrogen modulates intracellular calcium dynamics in smooth muscle cells.

Mathematical models of smooth muscle calcium dynamics have been developed across a range of applications, building on more general mathematical descriptions of intracellular calcium signalling~\cite{KeenerSneyd2008}. 
These models typically describe cytosolic Ca$^{2+}$ availability and its downstream influence on contractile force generation, and have been reviewed for different smooth muscle types in~\cite{garrett2022, leesgreen2011, tsoukias2011}. 
Cytosolic Ca$^{2+}$ concentration is governed by the coordinated processes of Ca$^{2+}$ influx, release from intracellular stores, sequestration, and extrusion. 
Calcium enters the cytosol primarily through voltage-dependent, receptor-operated, and store-operated channels, while release from the sarcoplasmic reticulum (SR) is mediated by IP$_3$ and ryanodine receptors. 
Elevated cytosolic Ca$^{2+}$ activates myosin light chain kinase via calmodulin, promoting actin--myosin cross-bridge cycling and contraction, whereas relaxation is achieved through Ca$^{2+}$ removal by SERCA, PMCA, and the Na$^+$/Ca$^{2+}$ exchanger (NCX)~\cite{karaki, somlyo, wray}. 
Many of these transport pathways have been shown experimentally to be regulated by estrogen signalling, suggesting that hormone-dependent modulation of intracellular calcium dynamics may play an important role in smooth muscle function.

In this work, we extend the established smooth muscle Ca$^{2+}$ dynamics model of Wang et al.~\cite{Wang2010}, chosen for its simplicity while retaining the principal mechanisms governing intracellular calcium handling, to incorporate estrogen receptor $\alpha$ (ER$\alpha$)-mediated modulation of calcium transport~\cite{Traupe2007}. 
To the best of our knowledge, this is the first mathematical model to explicitly represent estrogen signalling within a smooth muscle calcium dynamics framework. 
We then employ numerical simulation, global sensitivity analysis, and numerical bifurcation analysis to quantify how estrogen alters intracellular calcium dynamics, steady states, and oscillatory behaviour. 
As such, this work provides an initial mathematical framework for investigating hormone-dependent regulation of smooth muscle calcium signalling and establishes a foundation for future, more mechanistic models of estrogen-mediated smooth muscle physiology.

\section{Methods} 

\begin{figure}
\begin{center}
\begin{tikzpicture}[
    font=\small,
    flux/.style={-{Triangle[length=2.4mm,width=2mm]}, line width=1pt},
    estr/.style={-{Triangle[length=2.4mm,width=2mm]},
                 line width=1pt, red!65!black},
    dashmodarr/.style={dashed, -{Triangle[length=2mm,width=1.6mm]},
                       line width=0.8pt, red!60!black},      
    dashinhib/.style={dashed, -{Bar[width=4mm]},
                      line width=0.9pt, red!60!black},        
    gate/.style={draw=black, fill=white, line width=0.7pt,
                 minimum width=3mm, minimum height=7.5mm, inner sep=0pt},
    lbl/.style={inner sep=1.5pt},
  ]

  \draw[rounded corners=13pt, line width=1.1pt, fill=blue!4]
        (-4.2,-2.4) rectangle (4.2,2.4);
\node[lbl, font=\fontsize{12}{14}\selectfont] at (-1.4,1.2) {$c_{\mathrm{in}}$};
  \draw[line width=1.1pt, fill=orange!12]
        (1.3,-0.95) ellipse [x radius=1.9, y radius=1.05];
  \node[lbl, font=\fontsize{12}{14}\selectfont] at (1.3,-0.95) {$c_{\mathrm{SR}}$};

  \node[gate] (gin) at (-2.0,2.4) {};
  \draw[flux] (-2.0,3.0) -- (-2.0,1.70);
  \node[lbl] at (-2.7,2.1) {$J_{\mathrm{in}}$};
  \node[gate] (gpm) at (1.2,2.4) {};
  \draw[flux] (1.2,1.8) -- (1.2,3.1);
  \node[lbl] at (2,2.1) {$J_{\mathrm{PMCA}}$};

  \node[gate] (gse) at (-0.15,-0.3) {};
  \draw[flux] (-0.15,0.3) -- (-0.15,-1.03);
  \node[lbl] at (-0.9,0.3) {$J_{\mathrm{SERCA}}$};
  \node[gate] (gle) at (0.85,0.0) {};
  \draw[flux] (0.85,-0.5) -- (0.85,0.75);
  \node[lbl] at (0.43,0.9) {$J_{\mathrm{leak}}$};
  \node[gate] (gip) at (1.6,0.0) {};
  \draw[flux] (1.6,-0.5) -- (1.6,0.75);
  \node[lbl] at (1.65,1.05) {$J_{\mathrm{IP3R}}$};
  \node[gate] (gry) at (2.45,-0.25) {};
  \draw[flux] (2.45,-0.75) -- (2.45,0.5);
  \node[lbl] at (2.75,0.7) {$J_{\mathrm{RyR}}$};

  \node[draw, circle, line width=1pt, minimum size=9mm, fill=red!12] (E)
        at (-5.7,-0.3) {$E$};
  \node[gate, minimum width=7mm, minimum height=4mm] (er1) at (-4.2, 0.05) {};
  \node[gate, minimum width=7mm, minimum height=4mm] (er2) at (-4.2,-0.65) {};
  \node[lbl] (era) at (-3.3,-0.3) {$ER\alpha$};
  \draw[estr] (E.east) -- (-4.55,-0.3);

  \draw[dashinhib] (-3.3,-0.1) to[out=80,in=250] (-2.0,1.65);
  \draw[dashmodarr] (-2.85,-0.3) to[out=0,in=205] (-0.35,-0.3);
  \node[lbl, align=center, font=\fontsize{8}{0}\selectfont] at (1.3,-1.65) {Sarcoplasmic\\reticulum};
    \node[lbl, align=center, font=\fontsize{8}{0}\selectfont] at (0,-2.21) {Cytosol};
\end{tikzpicture}
\caption{A schematic representation of the key processes and channels governing Ca$^{2+}$ transport between the sarcoplasmic reticulum (SR), the cytosol and external to the cell. There is a flux into the cell, denoted by \Jin, and a flux out, denoted by \Jpmca. Between the cytosol and the SR, the flux denoted \Jserca  governs influx to the SR from the cytosol, whilst the fluxes \Jleak, \JIPthreeR and \Jryr are all outward fluxes allowing calcium to flow from the SR into the cytosol. These processes are in line with those described in~\cite{berridge2008smooth, Lukyanenko1996, Wang2010}, but the additional impact of ER$\alpha$ is shown in red to impact the flux into the cytosol and the flux into the SR.}
\label{fig:schematic}
\end{center}
\end{figure}
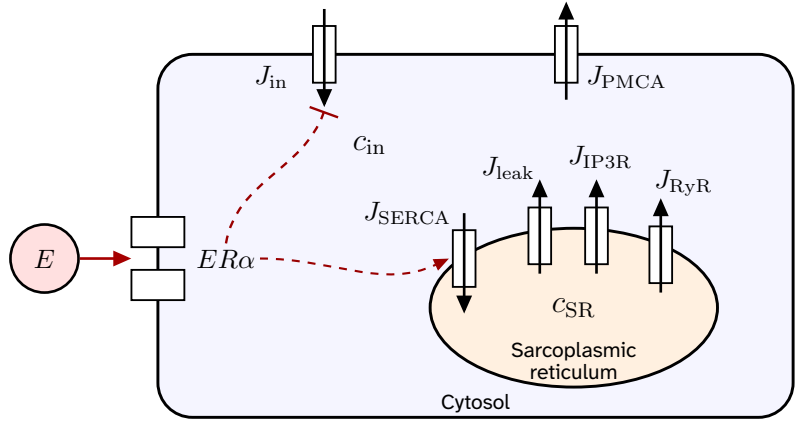

To consider the role of estrogen ($E$) in regulating smooth muscle cell Ca$^{2+}$ concentration, we extend the Wang model~\cite{Wang2010} of Ca$^{2+}$ signalling in airway smooth muscle cells to incorporate the effect of ER$\alpha$ activation. 
Unless otherwise stated, we retain the model parameterisation defined by Wang et al.~\cite{Wang2010}, acknowledging that parameterisation may need to be adjusted in future to represent different tissue types, but noting that the key signalling pathways exist across different smooth muscle types. Rather than modelling the complex signalling cascade initiated by ER$\alpha$ activation explicitly, estrogen is incorporated as a mechanistic representation of estrogenic signalling, introducing estrogen dependent scaling of the voltage dependent Ca$^{2+}$ current, denoted $I_{Ca}$, and of the sarcoplasmic reticulum (SR) Ca$^{2+}$ uptake. A schematic representation of the cell, and the fluxes considered in this work is shown in Fig.~\ref{fig:schematic}.

\subsection{Mathematical Model}
We consider a representative model of a single cell (as in Fig.~\ref{fig:schematic}) and focus only on the calcium dynamics moving between components in it. 
Whilst the extracellular calcium is considered to be in a constant, excess supply, the intracellular (cytosolic) Ca$^{2+}$ concentration, denoted $C_{\text{\text{in}}}$, and the SR Ca$^{2+}$ concentration, denoted $C_{\text{SR}}$, are governed by transmembrane and SR membrane ion channel activity. 
The following conservation equations describe the evolution of these concentrations in time
\begin{align}
    \frac{dC_{\text{\text{in}}}}{dt} &= J_{\text{\text{in}}} - J_{\text{SERCA}} + J_{\text{release}}  
        - J_{\text{PMCA}},\label{eq:C_in} \\
    \frac{dC_{\text{SR}}}{dt} &= \gamma\left(J_{\text{SERCA}} - J_{\text{release}}\right), 
       \label{eq:C_SR}
\end{align}
where $J_{\text{\text{in}}}$ is the voltage dependent transmembrane Ca$^{2+}$ influx, $J_{\text{SERCA}}$ is the SR Ca$^{2+}$-ATPase uptake flux, $J_{\text{release}} = J_{\text{IP3R}} + J_{\text{RyR}} + J_{\text{leak}}$ is the
total Ca$^{2+}$ flux from the SR to the cytosol, $J_{\text{PMCA}}$ is the plasma membrane Ca$^{2+}$-ATPase (PMCA) efflux, and $\gamma$ is a dimensionless geometric scaling factor equal to the cytosol-to-SR volume ratio~\cite{Wang2010}. 
Each flux term in Eqs.~\eqref{eq:C_in} and~\eqref{eq:C_SR} has its own governing equation. 
These are described in what follows, and their accompanying parameter values can be found in Table~\ref{tab:params}.

The voltage dependent transmembrane Ca$^{2+}$ influx is described by
\begin{align}
    J_{\text{\text{in}}} = \delta(\alpha_{0} + \alpha_{2}P - \alpha_{1} I_{Ca}) . \label{eq:Jin}
\end{align}
where $\alpha_{0}$ represents a constant basal Ca$^{2+}$ influx, $\alpha_2P$ represents the proportional IP$_3$ dependent store operated calcium influx (through agonist operated Ca$^{2+}$ channels (AOCC)), and the third term represents the proportional voltage dependent Ca$^{2+}$ current, $I_{Ca}$ (through voltage operated Ca$^{2+}$ channels (VOCC)) . 
Whilst $\alpha_{1}$ and $\alpha_{2}$ are proportionality constants, $\delta$ is a dimensionless parameter that scales between the transmembrane current and a cytosolic Ca$^{2+}$ flux.  
In what is considered here, the IP$_3$ concentration, represented by $P$, is assumed to be constant. 
The voltage dependent Ca$^{2+}$ current is described by 
\begin{align}
    I_{Ca} &= \frac{g_{\text{ca}}}{2F}m^{2}V_{Ca}{\color{black}\left(1 - \eta_{1}E \right)}, \label{eq:ICa}
\end{align}
where $g_{\text{ca}}$ is the conductance of Ca$^{2+}$ channels, $m$ is the voltage dependent activation gate, $V_{Ca}$ is the Ca$^{2+}$ reversal potential given by the Goldman-Hodgkin-Katz equation and $F$ is Faraday's constant~\cite{HaiMurphy1988, Wang2010}. 
The voltage dependent activation gate can be described by
\begin{align}
    m &=(1+e^{-(V-V_m)/k_m})^{-1}, \label{eq:m}
\end{align}
where $V$, $V_m$ and $k_m$ represent the voltage, the half-activation voltage and the activation slope, respectively.
Electrophysiological studies have shown that 17$\beta$-estradiol, a type of estrogen, acutely inhibits voltage-dependent L-type Ca$^{2+}$ currents in vascular smooth muscle cells~\cite{han1995, nakajima,tran2020,ueda2020}, that motivated the extension of the Wang model~\cite{Wang2008, Wang2010} to introduce an estrogen-dependent linear scaling term that reduces the voltage-dependent calcium current as estrogen concentration, denoted by $E$, increases.
$\eta_1$ is a proportionality constant, representing the strength of the estrogen effect. 

It has also been reported that long-term estrogen signalling influences intracellular Ca$^{2+}$ homeostasis through the regulation of SERCA channels and SR Ca$^{2+}$ handling, although the available experimental evidence is conflicting and the underlying molecular mechanisms remain incompletely understood~\cite{chu2006, jiao2020, nita2021, tran2020}. 
Given this uncertainty, we do not attempt to model the molecular pathways linking estrogen receptor activation to SERCA regulation explicitly. 
Instead, we phenomenologically extend the SERCA uptake formulation of Wang et al.~\cite{Wang2008, Wang2010} by introducing an estrogen-dependent scaling factor into the Hill function describing SERCA-mediated Ca$^{2+}$ uptake from the cytosol into the SR
\begin{align}
    J_{\text{SERCA}} = V_{s}\frac{C_{\text{\text{in}}}^{2}}{k_{s}^{2} + C_{\text{\text{in}}}^{2}}
        {\color{black}\left(1 + \eta_{2}E\right)}. \label{eq:Jserca}
\end{align}
In Eq.~\eqref{eq:Jserca}, $V_s$ is the maximal SERCA uptake rate, $k_s$ is the half maximal activation concentration, and $\eta_2$ is a scaling parameter quantifying the strength of the estrogen-dependent enhancement of SERCA-mediated Ca$^{2+}$ uptake. 
This approach captures the hypothesised enhancement of SR Ca$^{2+}$ sequestration while avoiding assumptions regarding the underlying regulatory mechanisms.

As previously described, SR Ca$^{2+}$ release occurs via three parallel pathways, each proportional to the concentration gradient between the SR and the cytosol, $C_{\text{SR}} - C_{\text{\text{in}}}$.
IP$_3$R mediated release is described by
\begin{align}
    J_{\text{IP3R}} = k_{\text{IP3R}}\,P_{\text{IP3R}}\left(C_{\text{SR}} - C_{\text{\text{in}}}\right),
        \label{eq:Jip3r}
\end{align}
where $k_{\text{IP3R}}$ is the maximal IP$_3$R permeability and $P_{\text{IP3R}}$ is
the channel opening probability given by
\begin{align}
    P_{\text{IP3R}} &= \left(\frac{P\,C_{\text{\text{in}}}(1-y_g)}{(P+K_1)(C_{\text{\text{in}}}+K_5)}
        \right)^{3}. \label{eq:P_IP3R} 
\end{align}
Eq.~\eqref{eq:P_IP3R} depends on the inositol 1,4,5-trisphosphate concentration, denoted $P = [\text{IP}_3]$, as well as the fraction of IP$_3$R inactivated by Ca$^{2+}$, denoted $y_g$ and itself governed by the following differential equation
\begin{align}
    \frac{dy_g}{dt} &= f_1(1 - y_g) - f_2\,y_g, \label{eq:dy} \\
    f_1 &= \frac{(k_{-4}K_2 K_1 + k_{-2}K_4\,P)\,C_{\text{\text{in}}}}{K_4 K_2(K_1 + P)},
        \label{eq:f1} \\
    f_2 &= \frac{k_{-2}\,P + k_{-4}K_3}{K_3 + P}. \label{eq:f2}
\end{align}
In Eqs.~\eqref{eq:f1} and~\eqref{eq:f2}, the parameters $k_i$ for $i=\{-5, -4, -3, -2, -1, 1, 2, 3, 4, 5\}$ describe forward (positive subscripts) and reversal (negative subscripts) rate constants, whilst $K_i=k_{-i}/k_i$ describes the ratio between the two. 

RyR mediated release is described by
\begin{align}
    J_{\text{RyR}} = k_{\text{RyR}}\,P_{\text{RyR}}\left(C_{\text{SR}} - C_{\text{\text{in}}}\right),
        \label{eq:Jryr}
\end{align}
where $k_{\text{RyR}}$ is the maximal RyR permeability and $P_{\text{RyR}}$ is the
channel open probability, given by 
\begin{align}
    P_{\text{RyR}}     &= \left(k_{\text{ryr0}} + \frac{k_{\text{ryr1}}\,C_{\text{\text{in}}}^{3}}
        {k_{\text{ryr2}}^{3} + C_{\text{\text{in}}}^{3}}\right)
        \frac{C_{\text{SR}}^{4}}{k_{\text{ryr3}}^{4} + C_{\text{SR}}^{4}},
        \label{eq:P_RyR}
\end{align}
where the Hill-3 term in $C_{\text{\text{in}}}$ captures cytosolic Ca$^{2+}$ activation via calcium induced calcium release (CICR), whilst the Hill-4 term in $C_{\text{SR}}$ models store dependent activation, reflecting the increase in RyR open probability during SR Ca$^{2+}$ overload~\cite{Shannon2004, Lukyanenko1996}. The descriptions of the remaining parameters, describing opening and activation rates, can be found in Table~\ref{tab:params}.

Finally, a passive SR leak is modelled as
\begin{align}
    J_{\text{leak}} = k_{\text{leak}}\left(C_{\text{SR}} - C_{\text{\text{in}}}\right),
        \label{eq:Jleak}
\end{align}
where $k_{\text{leak}}$ is a constant leak permeability, and PMCA mediated Ca$^{2+}$ efflux across the plasma membrane is modelled as a saturating Hill function of $C_{\text{\text{in}}}$, given by
\begin{align}
    J_{\text{PMCA}} = V_{p}\delta\frac{C_{\text{\text{in}}}^{n}}{K_{p}^{n} + C_{\text{\text{in}}}^{n}},
        \label{eq:Jpmca}
\end{align}
where $V_p$ is the maximal PMCA extrusion rate, $K_p$ is the half maximal activation concentration, and $n$ is the Hill coefficient, again all described in Table~\ref{tab:params}. 

\subsubsection{Estrogen levels}\label{sec:E}

Physiological estrogen concentrations vary over timescales of hours to weeks, whereas intracellular Ca$^{2+}$ oscillations and smooth muscle contraction occur over timescales of seconds.  We therefore exploit this separation of timescales by treating estrogen concentration as constant during each simulation. 
The parameter $E$ represents the prevailing level of estrogenic signalling and modulates the estrogen-dependent calcium transport pathways described above. Throughout this work we investigate how varying $E$ across its physiological range alters intracellular calcium dynamics while assuming that estrogenic signalling remains constant over the duration of each simulation.

Following the physiological ranges reported by Verdonk et al.~\cite{verdonk2019}, we consider representative circulating estradiol concentrations spanning approximately 7--700~pg\,mL$^{-1}$. These values encompass low-estrogen conditions, representative of males and post-menopausal women, as well as elevated estrogen concentrations observed during the late follicular phase of the menstrual cycle and early pregnancy~\cite{harrison1980, shaaban1973}.  Unless otherwise stated results are shown with $E=$7~pg\,mL$^{-1}$

\begin{center}
\begin{table}[htbp]
\centering
\caption{Parameter values for the Ca$^{2+}$ flux model. Parameters are taken from Wang et al.~\cite{Wang2010} and the associated CellML implementation unless otherwise stated.}
\label{tab:params}
\begin{tabular}{llllll}
\hline
\textbf{Flux} & \textbf{Parameter} & \textbf{Symbol} & \textbf{Value} 
    & \textbf{Units} & \textbf{Reference} \\
\hline
\multirow{3}{*}{$J_{\text{\text{in}}}$}
    & Basal influx rate          & $\alpha_0$  & 0.05   & $\mu$M\,s$^{-1}$        & \cite{Wang2010} \\
    & VOCC scaling               & $\alpha_1$  & 0.25   & dimensionless        & \cite{Wang2010} \\
    & AAOC scaling               & $\alpha_2$  & 1.0    & s$^{-1}$             & \cite{Wang2010} \\
    & Ca$^{2+}$ channel conductance & $g_{\text{Ca}}$ & 9.0 & nS\,$\mu$M$^{-1}$  & \cite{Wang2010} \\
    & Half-activation voltage    & $V_m$       & $-50$  & mV                   & \cite{Wang2010} \\
    & Activation slope           & $k_m$       & 12     & mV                   & \cite{Wang2010} \\
\hline
\multirow{3}{*}{$J_{\text{PMCA}}$}
    & Maximum pump rate          & $V_p$       & 4.5    & $\mu$M\,s$^{-1}$        & \cite{Wang2010} \\
    & Half-activation constant   & $K_p$       & 0.4    & $\mu$M                   & \cite{Wang2010} \\
    & Hill coefficient           & $n$         & 4      & dimensionless        & \cite{Wang2010} \\
\hline
\multirow{2}{*}{$J_{\text{SERCA}}$}
    & Maximum pump rate          & $V_s$       & 4.5    & $\mu$M\,s$^{-1}$        & \cite{Lytton1992, Wang2010} \\
    & Half-activation constant   & $k_s$       & 0.1    & $\mu$M                   & \cite{Lytton1992, Wang2010} \\
\hline
\multirow{6}{*}{$J_{\text{IP$_3$R}}$}
    & Channel density            & $k_{\text{IP$_3$R}}$ & 5.55 & s$^{-1}$    & \cite{Wang2010} \\
    & Forward rate constant      & $k_1$       & 2000   & $\mu$M$^{-1}$\,s$^{-1}$ & \cite{DeYoung1992, Wang2010} \\
    & Reverse rate constant      & $k_{-1}$    & 260    & s$^{-1}$             & \cite{DeYoung1992, Wang2010} \\
    & Forward rate constant      & $k_2$       & 1.0    & $\mu$M$^{-1}$\,s$^{-1}$ & \cite{DeYoung1992, Wang2010} \\
    & Reverse rate constant      & $k_{-2}$    & 1.05   & s$^{-1}$             & \cite{DeYoung1992, Wang2010} \\
    & Forward rate constant      & $k_3$       & 2000   & $\mu$M$^{-1}$\,s$^{-1}$ & \cite{DeYoung1992, Wang2010} \\
    & Reverse rate constant      & $k_{-3}$    & 1886   & s$^{-1}$             & \cite{DeYoung1992, Wang2010} \\
    & Forward rate constant      & $k_4$       & 1.0    & $\mu$M$^{-1}$\,s$^{-1}$ & \cite{DeYoung1992, Wang2010} \\
    & Reverse rate constant      & $k_{-4}$    & 0.145  & s$^{-1}$             & \cite{DeYoung1992, Wang2010} \\
    & Forward rate constant      & $k_5$       & 100    & $\mu$M$^{-1}$\,s$^{-1}$ & \cite{DeYoung1992, Wang2010} \\
    & Reverse rate constant      & $k_{-5}$    & 8.2    & s$^{-1}$             & \cite{DeYoung1992, Wang2010} \\
\hline
\multirow{5}{*}{$J_{\text{RyR}}$}
    & Channel density            & $k_{\text{RyR}}$  & 5.0    & s$^{-1}$      & \cite{Wang2010} \\
    & Basal open probability     & $k_{\text{ryr0}}$ & 0.0072 & s$^{-1}$      & \cite{Friel1995, Wang2010} \\
    & CICR activation rate       & $k_{\text{ryr1}}$ & 0.334  & s$^{-1}$      & \cite{Friel1995, Shannon2004, Wang2010} \\
    & CICR activation affinity   & $k_{\text{ryr2}}$ & 0.5    & $\mu$M            & \cite{Friel1995, Shannon2004, Wang2010} \\
    & SR load affinity           & $k_{\text{ryr3}}$ & 38     & $\mu$M            & \cite{Shannon2004, Wang2010} \\
\hline
$J_{\text{leak}}$
    & Passive leak rate          & $k_\text{leak}$    & 0.1    & s$^{-1}$             & \cite{Wang2010} \\
\hline
\multirow{2}{*}{Scaling}
    & SR-to-cytosol volume ratio & $\gamma$    & 5.5    & dimensionless        & \cite{Wang2010} \\
    & Transmembrane flux scaling & $\delta$    & 0.05   & dimensionless        & \cite{Wang2010} \\
    & Estrogen non-dimensionalisation 1 & $\eta_1$    &$0.00142$  &    ml/pg    & Assumed \\
    & Estrogen non-dimensionalisation 2 & $\eta_2$    &$0.00035$ &     ml/pg  & Assumed \\
\hline
\multirow{5}{*}{Initial conditions}
    & Cytosolic Ca$^{2+}$        & $C_{\text{\text{in}}}(0)$  & 0.112 & $\mu$M          & \cite{Wang2010} \\
    & SR Ca$^{2+}$               & $C_{\text{SR}}(0)$  & 24    & $\mu$M          & \cite{Wang2010} \\
    & IP$_3$R inactivation gate  & $y_g(0)$              & 0     & dimensionless & \cite{Wang2010} \\
    & Membrane potential         & $V(0)$              & $-60$ & mV          & \cite{Wang2010} \\
    & IP$_3$ concentration       & $P(0)$              & 0     & $\mu$M          & \cite{Wang2010} \\
\hline
\end{tabular}
\end{table}
\end{center}
\subsection{Numerical methods}
The governing equations~\eqref{eq:C_in}--\eqref{eq:Jpmca} describe a coupled system of nonlinear ordinary differential equations (ODEs) that describe the calcium dynamics in the cytosol and the SR, as well as the channel gating variables.  
No explicit analytical solutions to this model have yet been found, and thus we resort to a numerical study of the system's dynamics, implemented in Python using SciPy. All code used to produce the results in this work is available at \url{https://github.com/beckycrossley/Estrogen\_modelling}. 

Time-based simulations of the model were simulated over a finite time interval $[0,100]$ using numerical integration of the initial value problem prescribed by the governing ODEs and accompanying initial conditions in Table~\ref{tab:params}.
The initial conditions were obtained either from physiological resting values or from numerically computed steady states.
The ODE system was integrated using SciPy's \texttt{solve\_ivp} with adaptive step-size control.
Simulations were run until transient behaviour had decayed and the long-term dynamics of the system could be clearly identified. 
Time series, phase portraits, and calcium oscillation characteristics were then extracted from the resulting numerical solutions using a uniformly sub-sampled temporal grid for post-processing and visualisation.

Initially, we numerically identified the steady states of the system of equations~\eqref{eq:C_in}--\eqref{eq:Jpmca} using Newton's root-finding algorithm. 
To investigate the dependence of model behaviour on physiological parameters, numerical continuation techniques were employed to facilitate a bifurcation analysis. 
At each stage, solutions obtained from neighbouring parameter values were used as initial guesses to improve convergence and facilitate continuation along equilibrium branches.
The stability of each equilibrium was then determined from the eigenvalues of the Jacobian matrix evaluated at the steady state.
An equilibrium was classified as stable when all eigenvalues possessed negative real parts and unstable otherwise.

Furthermore, we employed the SALib v1.4.5 package~\cite{herman2017, iwanaga2022} in Python v3.9.1 to perform a global sensitivity analysis using the extended Fourier amplitude sensitivity test (eFAST)~\cite{saltelli1999} to assess the sensitivity of model outputs to uncertainty in the model parameters. 
Six search curves were used, each comprising $5000$ samples, and each curve was resampled $10\,000$ times, with the confidence interval set at $95\%$.
The outputted first-order sensitivity index, $S_i$, quantifies the fraction of output variance attributable to variation in parameter $i$ alone, when all other parameters are held fixed, whereas the total-order sensitivity index, denoted $S_{T,i}$, quantifies the variance attributable to parameter $i$ together with its interactions with all other parameters~\cite{marino2008}. 
Together, these indices identify which parameters most strongly influence model behaviour. 
Given that time-series simulations showed oscillatory solutions (see Fig.~\ref{fig:time-series}, for example), we considered the mean concentration, as well as the frequency and amplitude of the intracellular Ca$^{2+}$ oscillations in the sensitivity analysis. 
eFAST assigns each parameter a distinct sampling frequency, and the reliability of the resulting sensitivity estimates degrades as interference between these frequencies increases with the number of parameters analysed simultaneously.
As such, to produce clearer conclusions from our full parameter set, we partitioned this into two groups, where a separate eFAST analysis was run on each.  
The parameters identified as most influential in each group, ranked by total-order index, were pooled into a combined set of eight parameters. 
Sensitivity indices were then recomputed for this reduced set, together with a dummy parameter, $d$.

\section{Results}

In this section, we numerically explore the resulting dynamics of the system of equations~\eqref{eq:C_in}--\eqref{eq:Jpmca}.
To begin with, we examine the time-dependent solutions to the full model.

\begin{figure}[htbp]
    \centering
    \includegraphics[width=\linewidth]{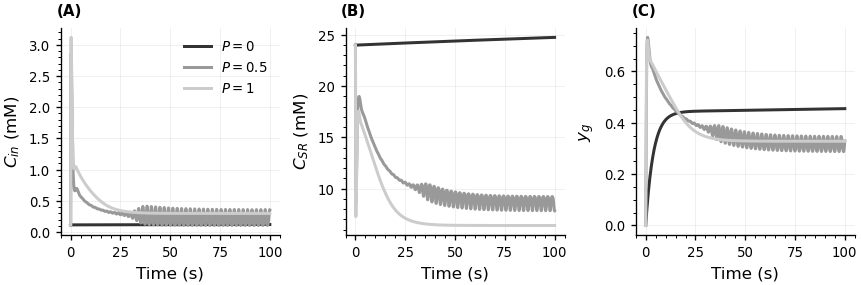}
    \caption{Time-dependent simulations of the calcium transport model described by Eqs.~\eqref{eq:C_in}--\eqref{eq:Jpmca} with baseline estrogen levels for three values of IP$_3$ agonist concentration ($P=0,\,0.5$, and $1$). Panels show (A) cytosolic calcium concentration, $C_\text{\text{in}}$, (B) sarcoplasmic reticulum (SR) calcium concentration, $C_\text{SR}$, and (C) the IP$_3$ receptor inactivation variable, $y_g$. }
    \label{fig:time-series}
\end{figure}

Fig.~\ref{fig:time-series} shows the numerical solutions of the model over time. 
For $P=0$, we find that solutions of the differential equations all converge to a stable steady state, and no oscillations can be observed. 
The calcium concentration in the SR increases with time as this model considers a constant external supply of calcium, which will continue to be drawn in. 
At intermediate concentrations ($P=0.5$), sustained oscillations in the calcium levels emerge following an initial transient period, which are well established by time $50$ and consistently oscillating within the same bounds by time $75$.
Interestingly, though, at larger concentrations, such as $P=1$, we find that the system once again approaches a stable equilibrium point over time, whose steady state values are lower than for smaller $P$ values. 
These results suggest that oscillations only occur in a finite range of $P$.

\subsection{Bifurcation analysis}
Motivated by the results in Fig.~\ref{fig:time-series}, that demonstrate the existence of stable steady state solutions in some regions of agonist parameter space, but oscillatory dynamics otherwise, we next investigated the existence of bifurcations in the system. 

\begin{figure}[htbp]
    \centering
    \includegraphics[width=\linewidth]{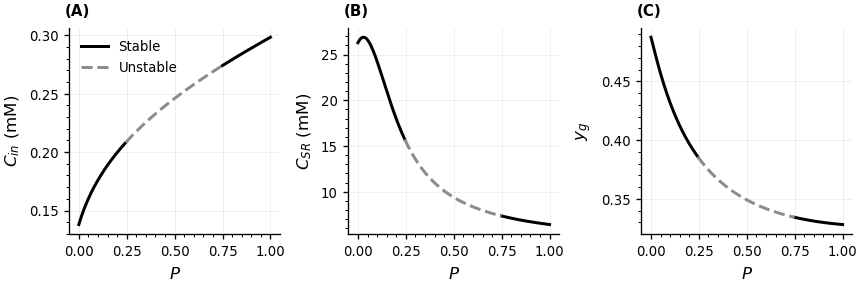}
    \caption{Equilibrium solutions with baseline estrogen levels of the calcium transport model (Eqs.~\eqref{eq:C_in}--\eqref{eq:Jpmca}) as the IP$_3$ agonist concentration parameter $P$ is varied. Panels show the steady-state values of (A) cytosolic calcium concentration, $C_\text{\text{in}}$, (B) sarcoplasmic reticulum (SR) calcium concentration, $C_\text{SR}$, and (C) the IP$_3$ receptor inactivation variable, $y_g$. Stable equilibria are shown by solid black curves and unstable equilibria by dashed grey curves.}
    \label{fig:bifurcation-P}
\end{figure}

Fig.~\ref{fig:bifurcation-P} shows the equilibrium solutions to Eqs.~\eqref{eq:C_in}--\eqref{eq:Jpmca} as the IP$_3$ agonist concentration parameter, $P$, is varied. 
We numerically find that a change in stability of the equilibrium steady state occurs through a Hopf bifurcation. 
The aforementioned change in stability is indicative of, and corresponds to, the onset of oscillatory calcium dynamics for intermediate values of $P$.
A later Hopf bifurcation at around $P=0.75$ shows that the system tends once again towards a stable steady state for larger $P$ values, and that no oscillations will be observed here.
These results are qualitatively similar to those found in previous studies by Wang et al.~\cite{Wang2010}, where estrogen was not included.

\begin{figure}[htbp]
    \centering
    \includegraphics[width=\linewidth]{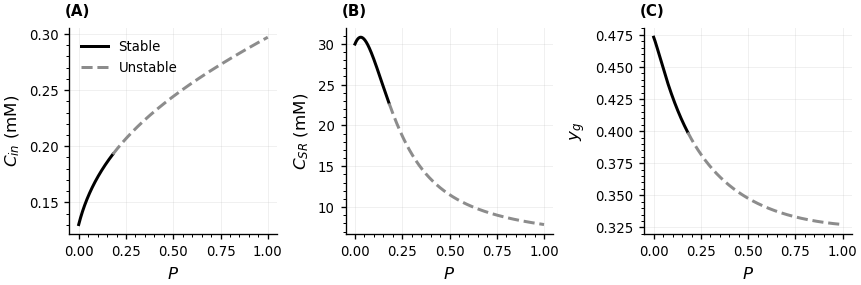}
    \caption{Equilibrium solutions with $E=$700~pg/,mL$^{-1}$ of the calcium transport model (Eqs.~\eqref{eq:C_in}--\eqref{eq:Jpmca}) as the IP$_3$ agonist concentration parameter $P$ is varied. Panels show the steady-state values of (A) cytosolic calcium concentration, $C_\text{\text{in}}$, (B) sarcoplasmic reticulum (SR) calcium concentration, $C_\text{SR}$, and (C) the IP$_3$ receptor inactivation variable, $y_g$. Stable equilibria are shown by solid black curves and unstable equilibria by dashed grey curves.}
    \label{fig:bifurcation-P-E700}
\end{figure}

Increasing $E$ to 700~pg/mL$^{-1}$expands the unstable region (see Fig.~\ref{fig:bifurcation-P-E700}) but with only small changes in the steady states. The 100-fold increase in $E$ compared to our baseline model shows qualitatively similar trends to the case of low estrogen, but the effect of increasing $E$ is to widen the unstable region and to slightly elevate the SR calcium concentrations at steady state.

\subsection{Sensitivity analysis}

\begin{figure}[htbp]
     \centering
        \begin{overpic}[width=\linewidth]{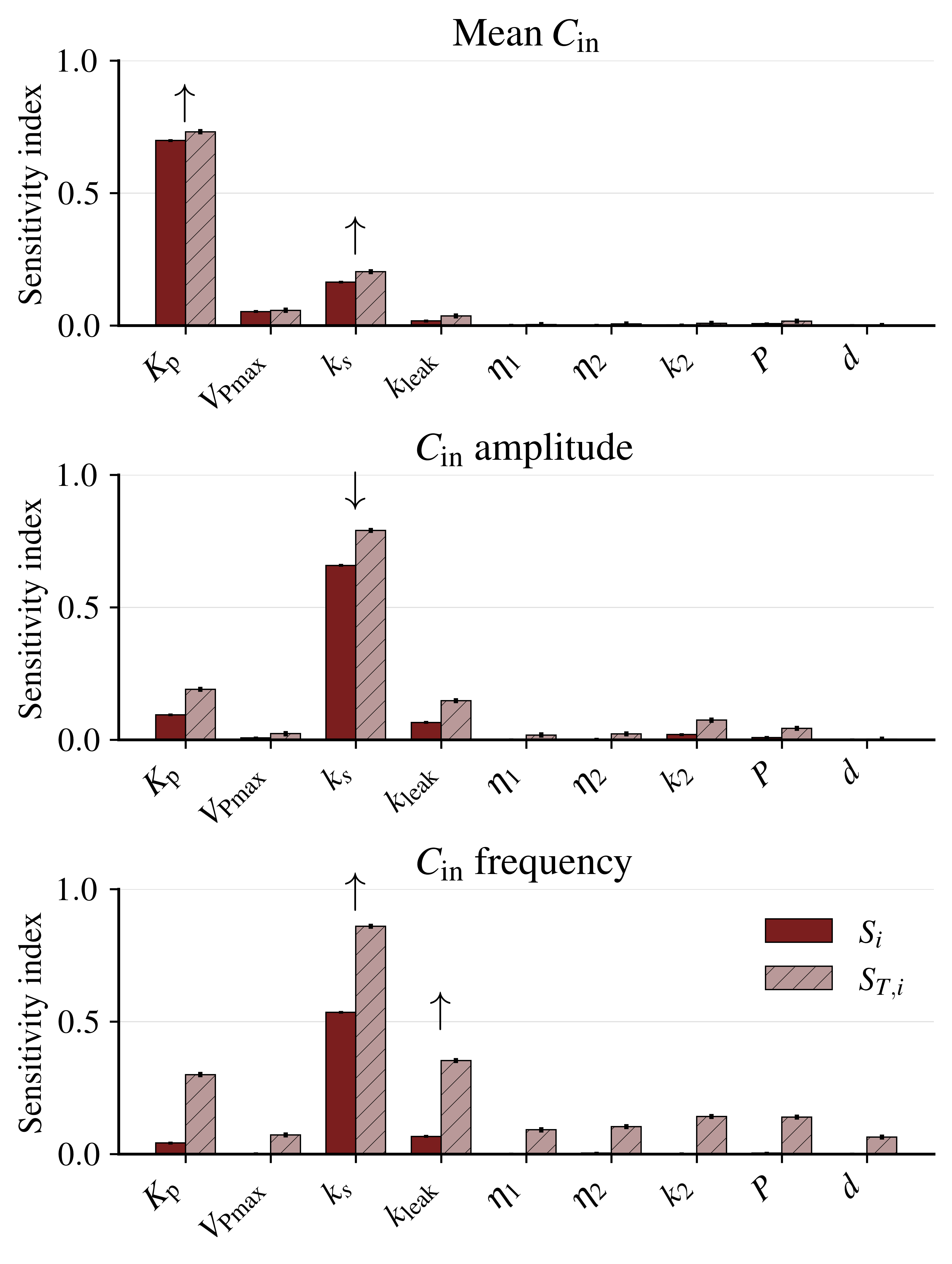}
        \put(0,97){\fontsize{9}{9}\selectfont {\bf (A)}}
         \put(0,65){\fontsize{9}{9}\selectfont {\bf (B)}}
         \put(0,33){\fontsize{9}{10}\selectfont {\bf (C)}}
        \end{overpic}
\caption{Results of sensitivity analysis showing the first-order (red) and total-order (pink) sensitivities for (A) mean $c_{\text{in}}$, (B) the $c_{\text{in}}$ amplitude, and (c) the $c_{\text{in}}$ frequency. The directions of influence for the most sensitive parameters are indicated by the black arrows above the bars. The label $d$ is the dummy variable. 
\label{fig:Sensitivity-Analysis}}
\end{figure}

Given that estrogen is known to affect not just smooth muscle cells in the airway, we wanted to investigate the impact of varying parameters in the model described by Eqs.~\eqref{eq:C_in}--\eqref{eq:Jpmca}. 
Fig.~\ref{fig:Sensitivity-Analysis} shows a bar chart of the eFAST sensitivity indices for each parameter, reporting both the first-order index $S_i$ (solid red) and the total-order index $S_{T,i}$ (hashed); the direction of each impactful parameter's effect on the output is indicated by an upward or downward arrow above the corresponding bar.

The results of this sensitivity analysis showed that the mean of $C_{\text{in}}$ is most sensitive to $K_{\text{P}}$, the half-saturation constant of PMCA, which increases the mean monotonically, with $k_s$ making a smaller, secondary contribution in the same direction. 
Comparatively, the remaining parameters show a marginal trend, with $\eta_1$ and $\eta_2$ in particular showing visually indistinguishable impacts from the dummy control. 
For both $K_{\text{P}}$ and $k_s$, $S_i$ and $S_{T,i}$ are closely matched, indicating that their influence on the mean is predominantly additive rather than interaction-driven.

eFAST also showed that the amplitude and frequency of the $C_{\text{in}}$ oscillatory dynamics are both most sensitive to $k_s$; the amplitude decreases monotonically with increasing $k_s$, whereas frequency increases. 
On the other hand, $K_{\text{P}}$ raises both amplitude and frequency. 
Although $k_2$ has only a modest first-order index, it shows a clear positive marginal relationship with frequency; this is consistent with its larger total-order index and indicates an appreciable interaction-mediated contribution.

Most notably, across all three metrics, the first-order indices of $\eta_1$ and $\eta_2$ are indistinguishable from that of the dummy parameter, indicating negligible direct (additive) sensitivity of the model to estrogen-mediated modulation of the L-type Ca$^{2+}$ current and SERCA flux. 
For oscillation frequency alone, both parameters have total-order indices modestly above the dummy, suggesting a weak effect mediated entirely through interactions with other parameters. 
Given the proximity of these values to the noise floor, this effect should be interpreted with caution, and would benefit from confirmation with a larger sample size or a targeted local sensitivity analysis.

\subsection{The role of changing estrogen levels in the cell}

\begin{figure}[htbp]
    \centering
    \includegraphics[width=\linewidth]{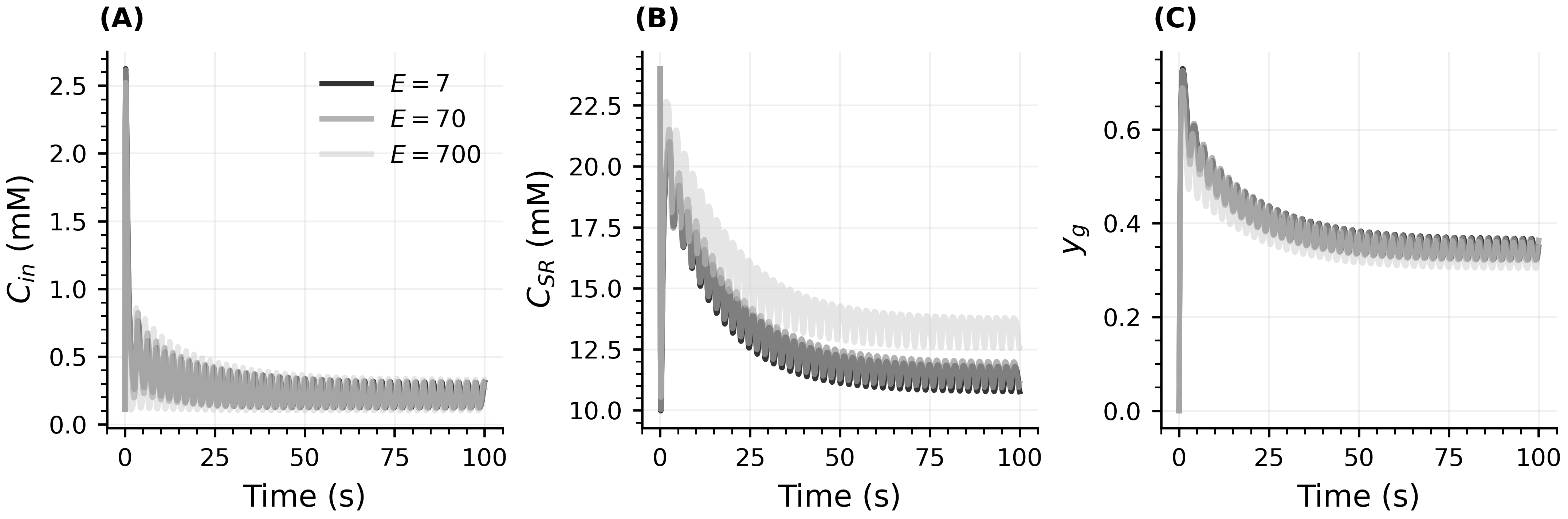}
    \caption{Time-dependent simulations of the calcium transport model described by Eqs.~\eqref{eq:C_in}--\eqref{eq:Jpmca} for three values of the estrogen concentration ($E=2.5,\,5$, and $25$). Panels show (A) cytosolic calcium concentration, $C_\text{\text{in}}$, (B) sarcoplasmic reticulum (SR) calcium concentration, $C_\text{SR}$, and (C) the IP$_3$ receptor inactivation variable, $y_g$.}
    \label{fig:time-E}
\end{figure}

Given the overall interest in the role of ER$\alpha$ on the intracellular calcium transport within a cell, we also wanted to investigate the role of changing the estrogen concentration in the environment of the cell. 
We began by considering the time series solutions of the calcium transport model described by Eqs.~\eqref{eq:C_in}--\eqref{eq:Jpmca} for varying estrogen levels, $E$ described in Sec.~\ref{sec:E}.
The results in Fig.~\ref{fig:time-E} clearly show that increasing the estrogen level increases the steady-state SR calcium concentration slightly, whilst the intracellular calcium levels are modulated only modestly via SERCA-mediated uptake. 
Whilst the size of the oscillations can also be observed to change slightly between estrogen concentrations, a bifurcation analysis (omitted for brevity) shows that, for the range of physiologically relevant parameters considered in this model, the Eqs.~\eqref{eq:C_in}--\eqref{eq:Jpmca} do not admit stable steady state solutions, when the default value for IP$_3$R concentration, $P=0.5$, is used. This aligns with the results shown in Figs.~\ref{fig:bifurcation-P} and~\ref{fig:bifurcation-P-E700}.

\section{Discussion}
Mathematical models have become valuable tools for investigating intracellular calcium signalling and identifying the nonlinear mechanisms responsible for observed oscillatory behaviours across a wide range of cell types~\cite{dupont2016, dupont, schuster}. 
Cytosolic Ca$^{2+}$ is crucial for smooth muscle cell contraction as it enables the mechanical machinery of the cell to perform its function. 
Although the influence of estrogen on smooth muscle physiology has been demonstrated experimentally~\cite{mendel1999, tran2020, white2010}, its effects have not previously been incorporated into mathematical models of smooth muscle calcium signalling. 
To our knowledge, this work therefore presents the first mathematical framework explicitly linking estrogen signalling to intracellular calcium dynamics in smooth muscle cells.

Our numerical simulations, sensitivity analysis, and bifurcation analysis demonstrate that estrogen acts primarily as a quantitative regulator of intracellular calcium dynamics. 
Increasing estrogen reduced cytosolic calcium concentrations and suppressed oscillatory behaviour through its modulation of calcium influx and SERCA-mediated uptake. 
However, despite these changes in calcium concentration and oscillation amplitude, the underlying qualitative dynamics of the model remained largely unchanged. 
In particular, the system retained the same bifurcation structure over the physiological range of estrogen considered, with estrogen shifting the location of critical transitions rather than introducing new dynamical regimes. 
These findings suggest that estrogen primarily modulates the operating point of the calcium signalling network rather than fundamentally altering its dynamical organisation.

The observed behaviour is consistent with current experimental understanding of estrogen signalling. 
Numerous studies report that estrogen reduces intracellular Ca$^{2+}$ availability through inhibition of L-type Ca$^{2+}$ channels together with activation of potassium channels that hyperpolarise the membrane and reduce calcium entry~\cite{han1995, hill2017, tran2020, valverde1999}. 
Such mechanisms would be expected to reduce smooth muscle contractility without necessarily changing the underlying regulatory architecture governing calcium oscillations. 
Our model provides quantitative support for this hypothesis by demonstrating how relatively simple modulation of calcium transport pathways can account for experimentally observed reductions in intracellular calcium while preserving the characteristic nonlinear dynamics of the signalling system.

One mechanism incorporated into the present model is estrogen-dependent enhancement of SERCA-mediated calcium uptake. 
Long-term estrogen signalling has been associated with increased SR Ca$^{2+}$ sequestration through altered SERCA expression; however, experimental evidence for this relationship remains conflicting, and the underlying molecular mechanisms are not yet fully understood~\cite{chu2006, jiao2020, nita2021, tran2020}. 
Several studies report restoration of SERCA expression following estrogen replacement in ovariectomised animal models~\cite{jiao2020}, whereas others observe little or no effect of estrogen on SERCA, instead identifying modulation of L-type Ca$^{2+}$ channels and the Na$^+$/Ca$^{2+}$ exchanger as the dominant mechanisms~\cite{chu2006}. 
Direct evidence for ER$\alpha$-mediated regulation of SERCA activity in vascular smooth muscle remains limited. 
Consequently, the present implementation should be regarded as a phenomenological representation of proposed estrogenic mechanisms rather than a definitive molecular description. 
As further experimental evidence becomes available, the model can readily be refined to incorporate more mechanistic descriptions of estrogen-dependent calcium transport.

The present study necessarily makes several simplifying assumptions. 
Estrogen signalling is represented by a single parameter acting on selected calcium transport pathways, while circulating estrogen concentration is assumed constant over the duration of each simulation because hormonal fluctuations occur over much longer timescales than intracellular calcium oscillations. 
Furthermore, the underlying Wang model does not explicitly represent extracellular calcium dynamics, membrane electrophysiology, downstream force generation, or receptor signalling pathways beyond those directly influencing calcium transport~\cite{gunst}. 
These simplifications allow the model to remain sufficiently low-dimensional for sensitivity and bifurcation analysis, while still capturing the principal mechanisms governing intracellular calcium dynamics.

There are several natural directions for future work. 
The model could be extended to include dynamic membrane voltage, extracellular calcium, or cross-bridge mechanics to provide a more complete description of smooth muscle contraction. There is certainly evidence that, as well as an influence on calcium dynamics, estrogen influences contractile force independently of calcium, for example, via an indirect nitric oxide mediated influence of companion cells~\cite{freay1997,Traupe2007}. Coupling the calcium signalling model to endocrine models describing cyclic estrogen variation would permit investigation of calcium dynamics across the menstrual cycle, pregnancy, and menopause.  Likewise, gradual reductions in estrogen could be used to investigate long-term changes in smooth muscle excitability associated with ageing and menopause, while simulated hormone replacement therapies could provide a quantitative framework for exploring how different treatment strategies modify intracellular calcium signalling and contractility~\cite{rossouw}. 
More broadly, the framework developed here provides a foundation upon which increasingly mechanistic models of estrogen-dependent smooth muscle physiology can be constructed as experimental understanding of hormone signalling continues to improve.

\paragraph{Acknowledgements}
RMC acknowledges support from the Engineering and Physical Sciences Research Council and MATRIX for travel funding that facilitated this research. 

\bigskip

\bibliographystyle{plain}
\bibliography{MatrixBibliography}

\end{document}